\documentclass[apj]{emulateapj}
\usepackage{graphicx}
\usepackage[bookmarks=true,colorlinks=true,citecolor=blue,linkcolor=magenta,urlcolor=cyan]{hyperref}
\usepackage{natbib}
\usepackage{amsmath}
\usepackage{amssymb}
\usepackage{subfigure}
\usepackage{url}
\usepackage{CJK}
\usepackage{float}
 
\def\Ha{H$\alpha$ \,} 
\def\Hb{H$\beta$ \,} 
\def\i{\,{\small I}}
\def\ii{\,{\small II}} 
\def\iii{\,{\small III}}

\slugcomment{Accepted for publication in ApJ}
\shorttitle{Building disk in IC~719} 
\shortauthors{Katkov et al.}

\begin{document}

\title{Lenticular galaxy IC 719: current building of the counterrotating
large-scale stellar disk.\footnote{Based on observations collected with the 6-m
telescope of the Special Astrophysical Observatory  of the Russian Academy of
Sciences which is operated under the financial support of Science Department of
Russia (registration number 01-43)}}

\author{Ivan Yu. Katkov\altaffilmark{1},
        Olga K. Sil'chenko\altaffilmark{1},
        Victor L. Afanasiev\altaffilmark{2}}
\email{katkov.ivan@gmail.com, olga@sai.msu.su, vafan@sao.ru}

\altaffiltext{1}{Sternberg Astronomical Institute, M.V.
Lomonosov Moscow State University, Moscow, 119992 Russia}
\altaffiltext{2}{Special Astrophysical Observatory,
Russian Academy of Sciences, Nizhnii Arkhyz,  Karachaevo-Cherkesskaya Republic,
369167 Russia}

\begin{abstract} 
We have obtained and analyzed long-slit spectral data for the lenticular galaxy
IC~719. In this gas-rich S0 galaxy, its large-scale gaseous disk counterrotates
the global stellar disk. Moreover in the IC~719 disk we have detected a
secondary stellar component corotating the ionized gas. By using emission-line
intensity ratios, we have proved the gas excitation by young stars and so are
claiming current star formation, most intense in a ring-like zone at the radius
of $10\arcsec$ (1.4 kpc). The oxygen abundance of the gas in the starforming
ring is about half of the solar abundance. Since the stellar disk remains
dynamically cool, we conclude that smooth prolonged accretion of the external
gas from a neighboring galaxy provides current building of the thin large-scale
stellar disk.

\end{abstract}


\section{Introduction}

Lenticular galaxies represent rather common morphological type in the nearby
Universe: they constitute up to 15\%\ of all galaxies in the field \citep{apm}
being at the second place after spirals.  Though they are more frequent in
clusters, there is also a considerable population of quite isolated lenticulars
\citep{amiga2}, or of lenticulars in isolated pairs and triplets. Lenticulars
are thought to lack gas in general; however just in the field the cold neutral
hydrogen is found in up to 25\% --45\%\ of all lenticulars, by the contrast
with the cluster members \citep{balk,alfalfa2,atlas3d_13}. The difference in
the gas detection frequency depending on environment density is sometimes
explained by numerous mechanisms of gas removing from the large-scale disks
which are effective only in dense environments. But this scenario cannot
explain the difference in gas kinematics depending on environment density: in
the cases when the gas is found in cluster lenticulars, its rotation matches
usually the rotation of the stellar components, while in the field S0s the gas
kinematics is very often decoupled from that of the stellar components
\citep{atlas3d_10}. Probably, the main difference related to the environments
must be looked for in the area of the gas acquisition conditions, and not in
the area of its subsequent dynamical evolution. \citet{atlas3d_10} suggest that
the gas in cluster lenticulars is provided by the mass loss of the stellar
components, and so comes from an intrinsic source. Field lenticulars are
evidently feeded by external accretion; but the sources of the external
accretion are not often recognized unambiguously. There are two main mechanisms
that have been suggested by theorists, and to make choice between them -- minor
mergers or cold accretion from large-scale filaments -- we need particular
observational investigations `in depth'. So every new S0 galaxy with a
significant kinematically decoupled gas component attracts close attention of
investigators.

Though small amounts of kinematically decoupled gas are common in lenticular
galaxies settling in low-density environments \citep{bertola}, large
counterrotating gaseous disks are rare, as well as gas-rich S0 galaxies at all.
In the (R)SA(rs)0+ galaxy NGC~3626 \citep{ciri,n3626mol,n3626hi}, located at
the periphery of a massive X-ray detected group, all the gas found, including
ionized, molecular, and neutral species, counterrotates the stellar component.
In unbarred galaxies and also group members, -- in SA0/a-galaxy NGC~3593
\citep{n3593} and in the SA(r)0+ galaxy NGC~4138 \citep{n4138thak} -- their
counterrotating gas is already partly processed into stars, so these galaxies
demonstrate two counterrotating stellar disks one of which rotates together
with their gas. In the SA0 NGC~1596 \citep{n1596} the outer gas, freshly
accreted from the neighboring NGC~1602, counterrotates the inner part of the
lenticular galaxy: the gas radial inflow in this unbarred galaxy is perhaps
ineffective. To this list we have recently added two more lenticular galaxies
with extended counterrotating gas, NGC~2551 and NGC~5631 \citep{ourcounter}; in
the former one the GALEX detects also a broad ring of star formation in the
disk-dominated area, in the latter we found two counterrotating stellar
components.  Interestingly, all the galaxies mentioned above are unbarred
though some of them possess rings and some reveal signatures of oval distortion
\citep{n4138_4550,leo2lent} which may be traces of past interaction.

\begin{table} 
\caption[ ] {Global parameters of the galaxies} 
\begin{tabular}{lc} 
\hline
\noalign{\smallskip}
Galaxy & IC 719  \\ 
Type (NED$^1$) & S0?\\ 
$R_{25}$, kpc (NED$+$RC3$^2$) &5.5\\ 
$B_T^0$ (LEDA$^3$) &  13.66 \\ 
$M_B$ (LEDA)  & --18.6  \\ 
$M_K$ (ATLAS-3D)  & --22.7  \\ 
$V_r $ (NED) & 1860 $\mbox{km} \cdot \mbox{s}^{-1}$\\ 
Distance, Mpc (ATLAS-3D)  & 29.4 \\ 
Inclination (LEDA) & $90^{\circ}$  \\
{\it PA}$_{phot}$ (LEDA)  &  $52^{\circ}$ \\ 
$V_{rot} \sin i$, $\mbox{km} \cdot\mbox{s}^{-1}$, (LEDA, HI) & $114.4 \pm 6.6$ \\ 
$\sigma _*$, $\mbox{km} \cdot\mbox{s}^{-1}$, (LEDA) & 121 \\ 
$M_{HI}$, $10^9\,M_{\odot}^4$ & 0.54 \\
$M_{H_2}$, $10^8\,M_{\odot}^5$ & 1.8 \\ 
\hline
\multicolumn{2}{l}{$^1$\rule{0pt}{11pt}\footnotesize NASA/IPAC Extragalactic
Database}\\ 
\multicolumn{2}{l}{$^2$\rule{0pt}{11pt}\footnotesize Third
Reference Catalogue of Bright Galaxies}\\
\multicolumn{2}{l}{$^3$\rule{0pt}{11pt}\footnotesize Lyon-Meudon Extragalactic
Database}\\ 
\multicolumn{2}{l}{$^4$\rule{0pt}{11pt}\footnotesize
\citet{alfalfa2,atlas3d_13}}\\
\multicolumn{2}{l}{$^5$\rule{0pt}{11pt}\footnotesize \citet{atlas3d_4}}\\
\end{tabular} 
\end{table}

\begin{figure*}[!t]
\epsscale{1} 
\plottwo{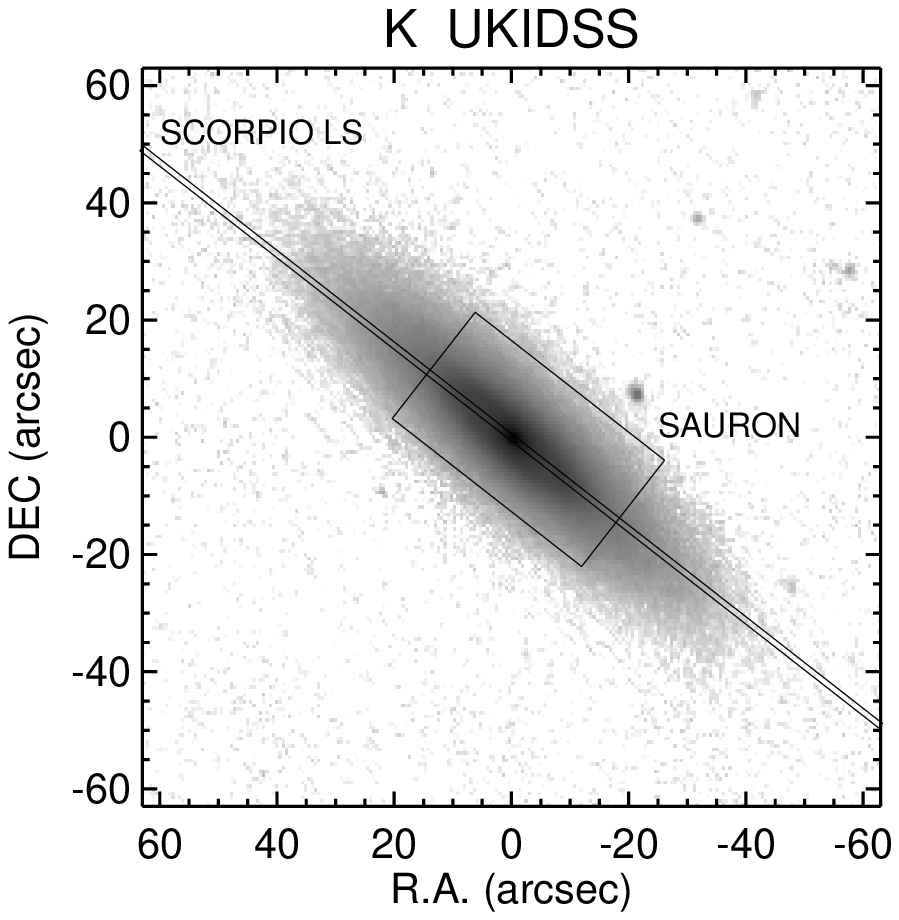}{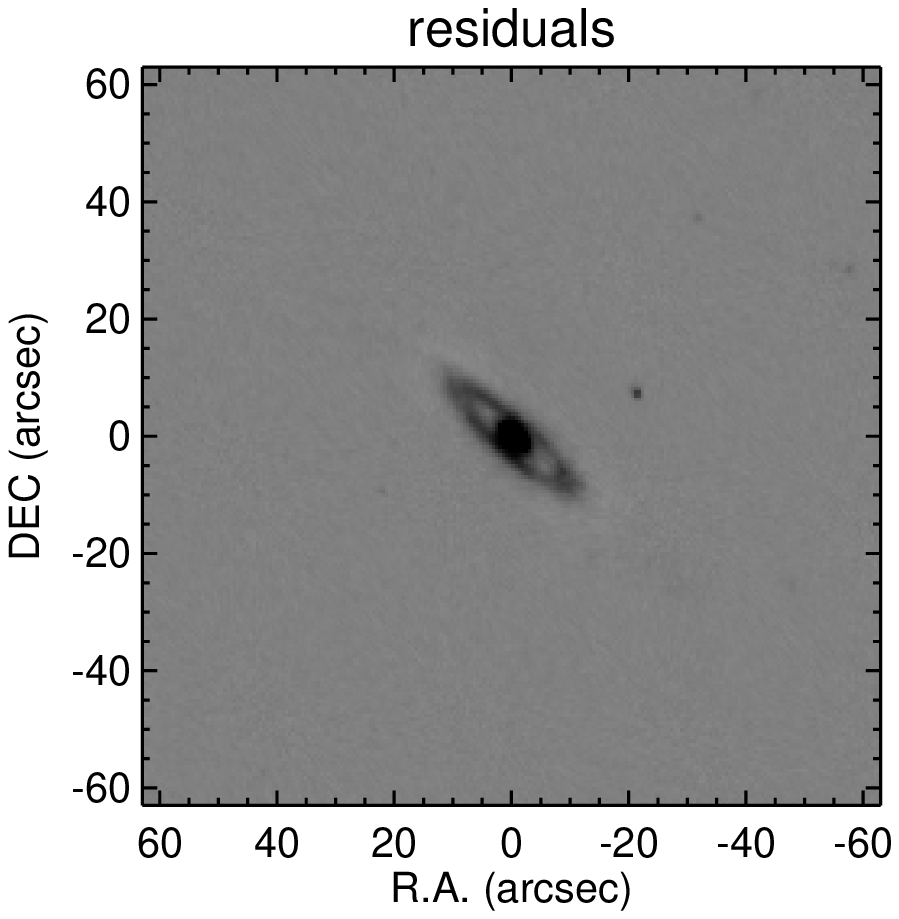}
\centering
\includegraphics[width=0.7\textwidth]{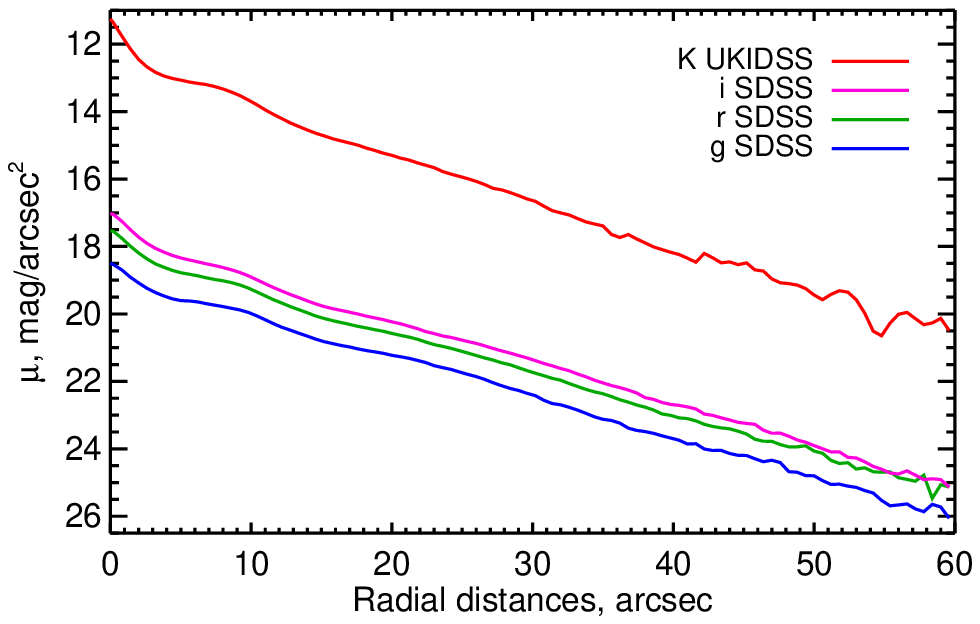}
\caption{S0 galaxy IC~719: top left -- a $K$-band image from the UKIDSS DR8;
top right -- residuals between $K$-band image and outward exponential disk
model; bottom -- the azimuthally averaged surface brightness profiles
calculated from the SDSS $g,r,i$-band images and the UKIDSS $K$-image.}
\label{view_prof} 
\end{figure*}

In the present paper, to the small list of the S0s with extended
counterrotating gas, and to even smaller list of S0s with two counterrotating
stellar components \citep{fisherkin}, we are adding one more interesting
object.  We have undertaken an analysis of various spectral data for the
lenticular galaxy IC~719. The galaxy is of intermediate luminosity and
constitutes an isolated non-interacting pair with IC~718 -- a late-type galaxy
of the similar luminosity.  General characteristics of IC~719 are presented in
the Table~1, and the general view can be seen in Fig.~\ref{view_prof}. Though
the galaxy is classified as an early-type one in all catalogues and databases,
one can assured that it is almost bulgeless: its surface brightness profiles
represent a pure quasi-exponential disk (Fig.~\ref{view_prof}). The galaxy is
rather gas-rich for the S0 type; moreover, neutral HI and molecular gas both
are detected. The galaxy was included into the sample of the survey ATLAS-3D
\citep{atlas3d_1} and observed with the integral-field spectrograph SAURON
\citep{sauron}; as a result, the counterrotation of the gaseous and stellar
components has been detected for its central part, $R<20\arcsec$.  Meantime,
the HI disk of IC~719 is very extended -- much more extended than even its
stellar disk \citep{alfalfa2}; however the large-scale velocity field of the HI
is unknown. It would be interesting to probe the sense of gas rotation beyond
the SAURON field of view, up to the optical border of the stellar disk -- it is
a sure chance to estimate the space scale of external gas accretion. By
measuring oxygen abundance in the extended gaseous disk, we can also make a
choice between various possible accretion sources. And the last, but not the
least aim of our analysis is a search for the secondary stellar component which
may be coupled with the counterrotating gas: the detailed star formation
history in the counterrotating gaseous disk is probably coupled with the
accretion history and so allows to reconstruct the accretion history and to
identify its source. We have fulfilled long-slit spectral observations of
IC~719 and are presenting now their results. Also we involve panoramic
spectroscopic data from the ATLAS-3D survey \citep{atlas3d_1} into our
analysis. The layout of the paper is the following: Section 2 describes our
observations and data reduction, Section 3 is dedicated to description of our
analysis approaches, in Section 4 we discuss the results on kinematics of the
stellar and gaseous components and their chemical compositions. In Section 5 we
conclude.

\section{Observations and data reduction} 

We used two datasets obtained from long-slit and integral-field spectroscopy.
The long-slit spectroscopic observations were made with a new universal
spectrograph SCORPIO-2 \citep{scorpio2} at the prime focus of the Russian 6-m
BTA telescope operated by the Special Astrophysical Observatory, Russian
Academy of Sciences. IC~719 was observed in November 2011 with the $1''$ slit
aligned along the major axis, with the total exposure times of $7 \times 1200$
sec.  The median atmosphere seeing FWHM during these observations was 1.5
arcsec.  We used the VPHG1200 grizm providing an intermediate spectral
resolution FWHM $\approx 3.5$ \AA\ in a wavelength region from 4300 to 7300
\AA.  This spectral range included a set of strong absorption and emission
features making it suitable to study both internal stellar and gaseous
kinematics and the stellar populations of the galaxy.  The slit was $6'$ in
length providing a possibility to use the edge spectra to evaluate the sky
background. The CCD chip E2V CCD42-90, with a format of $2048 \times 4600$,
using in the $1\times2$ binning mode provided spatial scale of 0.357 $''/$px
and a spectral sampling of 0.84 \AA/px.    

The preliminary data reduction was identical to that applied to the lenticular
galaxy NGC~7743 described in \citet{ngc7743}.  Briefly, the primary data
reduction comprised bias subtraction, flat-fielding, cosmic ray hit removal,
building the wavelength solution using the He-Ne-Ar arc-line spectra.  To
subtract the sky background, we invented a rather sophisticated approach.  We
constructed the spectral line spread function (LSF) model varied along and
across the wavelength direction by using the twilight spectrum
\citep{sky_subtr}. The final stages of the long-slit spectra reduction were
night sky spectrum subtraction taking into account the LSF variations,
linearization and accounting spectral sensitivity variation using the spectrum
of a spectrophotometrical standard star. The error frames were computed using
the photon statistics and processed through the same reduction steps as the
data.

We used also the data obtained with the integral-field spectrograph SAURON
\citep{sauron} mounted at the 4.2 m William Hershel Telescope, La Palma, in the
frame of the ATLAS-3D survey \citep{atlas3d_1}. The raw science and calibration
exposures were retrieved from the open Isaac Newton Group Archive of the
Cambridge Astronomical Data Center. The field of view of the spectrograph is
$44 \times 38$ spatial elements (spaxels) at a $0.94''$ scale per spaxel. The
SAURON spectral range is 4800-5350 \AA\ with spectral resolution FWHM of
$\approx 4$~\AA. For details of how we reduced the SAURON raw data see
\citet{SFH_sauron}.

\section{Internal kinematics and stellar populations} 

\subsection{{\sc nbursts} fitting} 
\label{nburst_description} 

\begin{figure}[!Hb]
\epsscale{1} 
\centerline{
\includegraphics[width=0.25\textwidth]{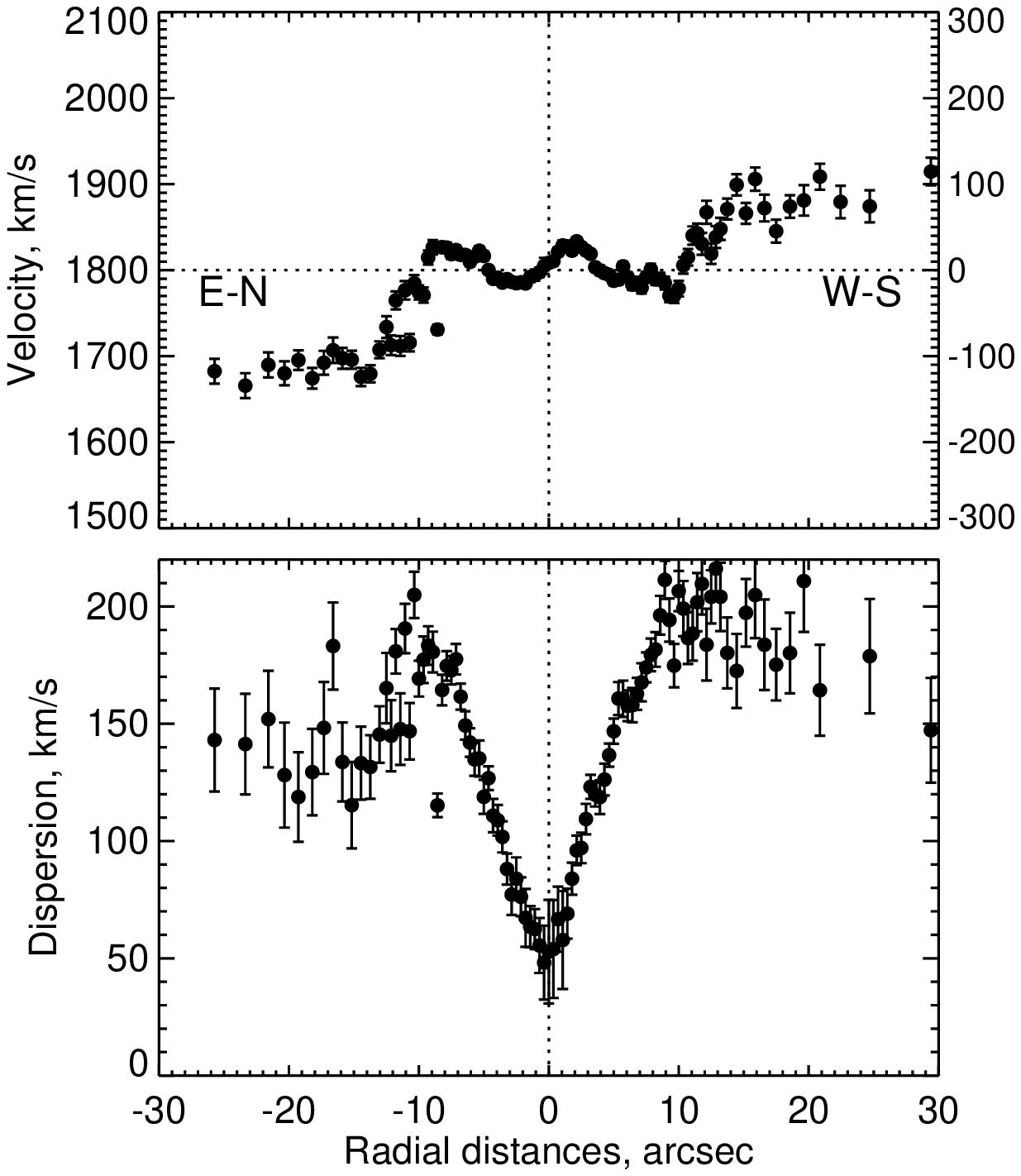}
\includegraphics[width=0.25\textwidth]{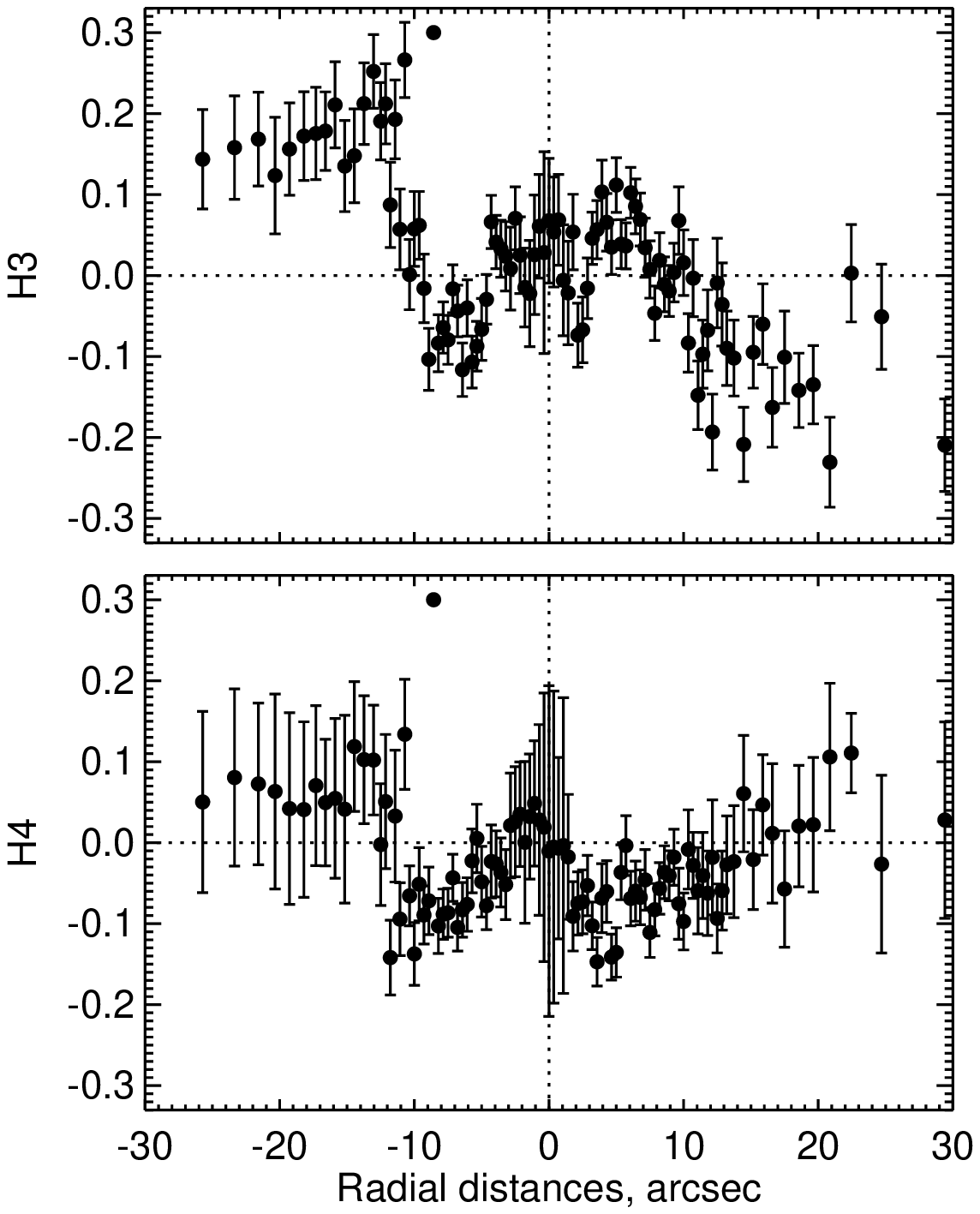}}
\caption{The results of the analysis of the long-slit spectrum for IC~719: left
top -- stellar LOS velocity profile; left bottom -- stellar velocity dispersion
profile, right top and bottom -- Hermite coefficients $h_3$, $h_4$.  The
binning is made along the slit to maintain $S/N=15$ per pixel. }
\label{ic719_kin_nbursts}
\end{figure}

\begin{figure*}[!Ht] 
\epsscale{1} 
\centerline{
\includegraphics[width=0.5\textwidth]{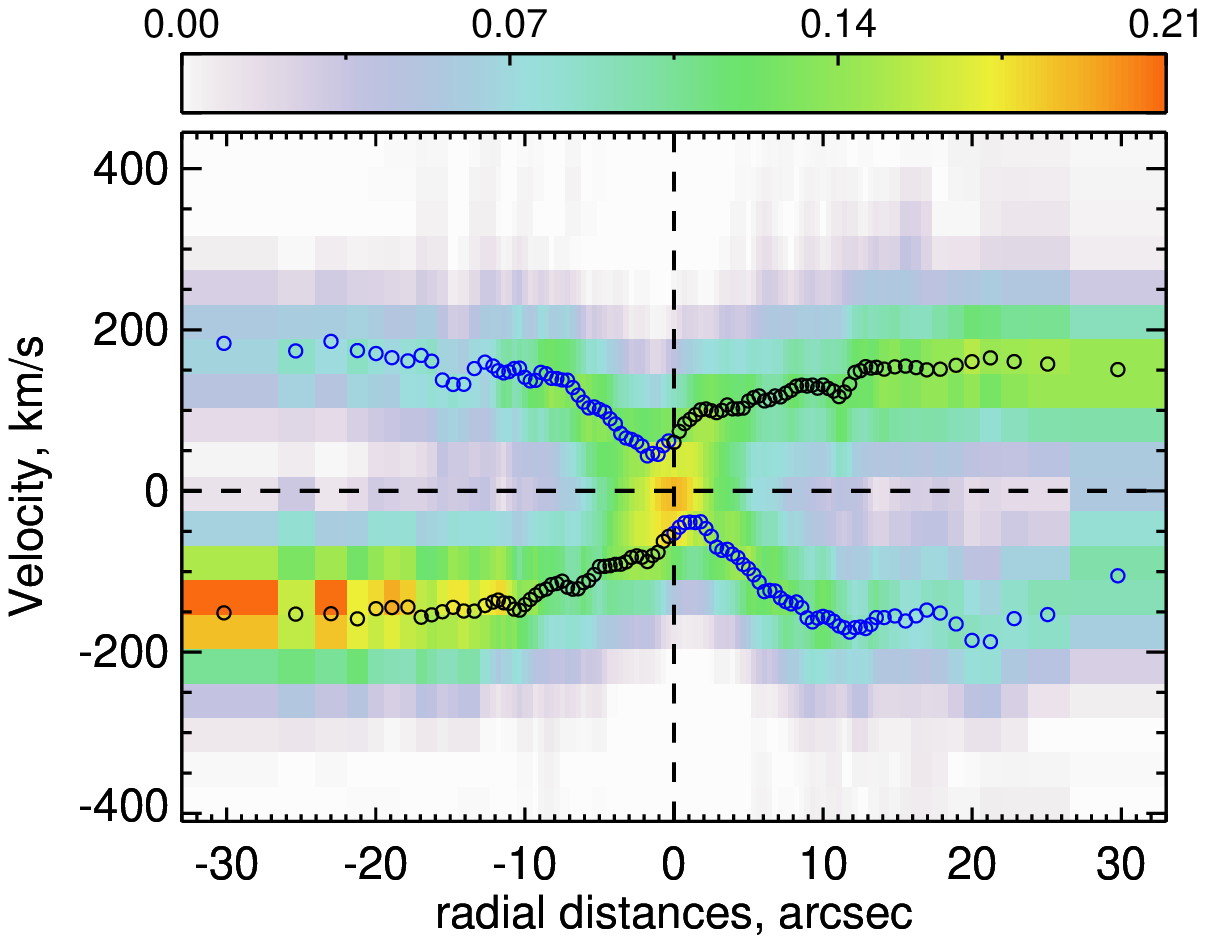}
\includegraphics[width=0.5\textwidth]{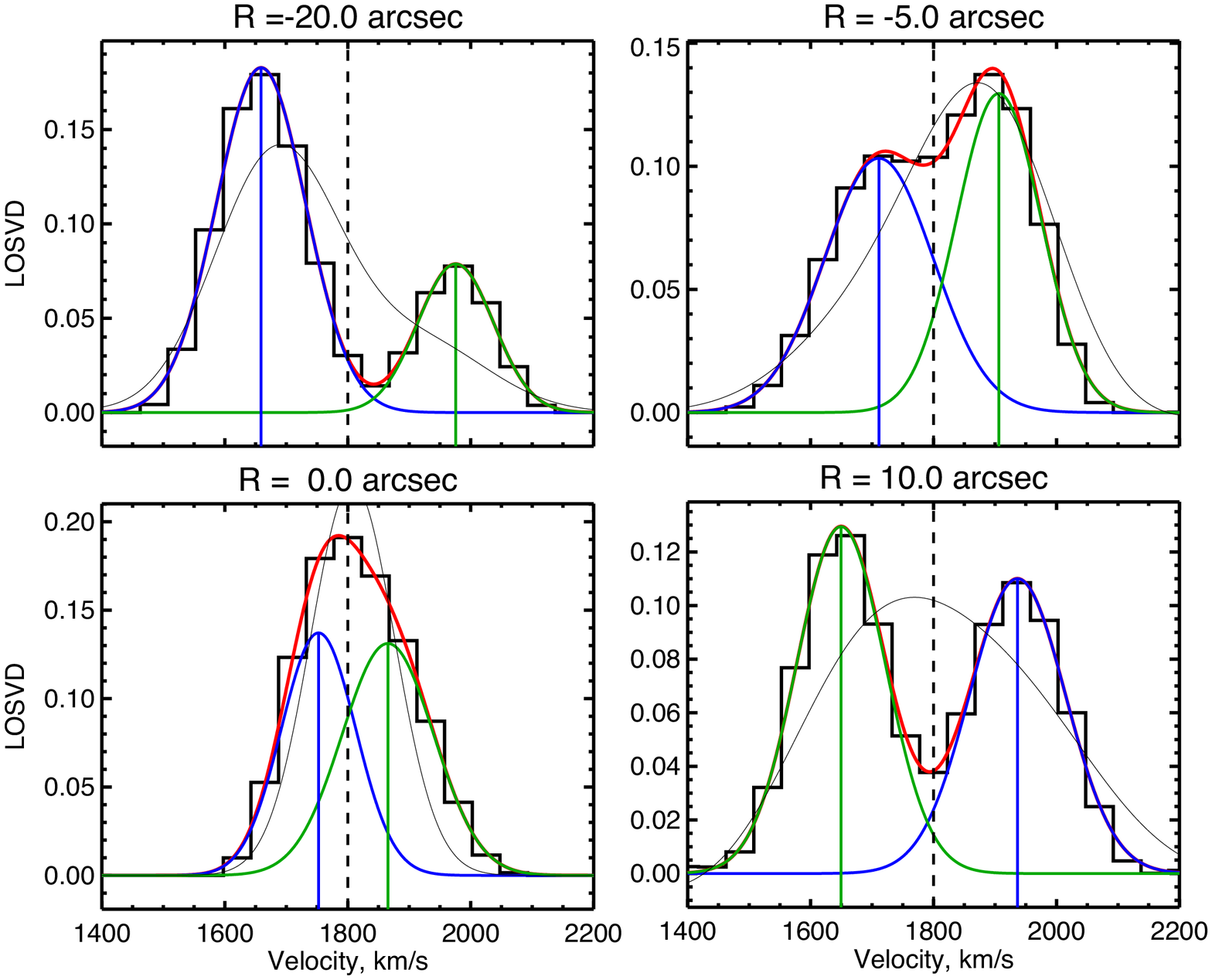}}
\caption{The view of the stellar LOSVD for IC~719 obtained by non-parametric
technique. Left -- stellar position-velocity diagram for the long-slit
spectrum. Black and blue circles mark position of gaussian centres for
two-gaussian decomposition. Right -- cross-sections of the PV diagram at the
different radial distances ($r=-20,-5,0,10$ arcsec). Solid black line presents
recovered stellar LOS at given distances, red line - model of the sum of two
separate gaussians (green and blue lines). Thin black lines correspond
to Gauss-Hermite approximation of the stellar LOS recovered under one-component
approach using the {\sc nbursts} fitting technique (see
Section~\ref{nburst_description}).} 
\label{PV_diag} 
\end{figure*}
 
We derived the parameters of internal stellar kinematics and stellar population
properties of IC~719 by fitting high-resolution PEGASE.HR \citep{pegase} single
(simple) stellar population (SSP) models, which were computed with the Salpeter
stellar initial mass function, to our spectra by using the \textsc{nbursts}
full spectral fitting technique \citep{nburst_a,nburst_b}. Before the main
minimization loop a grid of the SSP spectra with a fixed set of ages and
metallicities is convolved with the instrumental response of the spectrograph
(LSF). The non-linear least-square fitting algorithm is applied to select a
template spectrum from the grid of the SSP models by specifying its
luminosity-weighted (`SSP-equivalent') age $t$ and metallicity [Z/H]. Then we
convolve the template spectrum with a line-of-sight velocity distribution
(LOSVD) approximated by the Gauss-Hermite parametrization up to the $4^{th}$
moment, i.e. $v$, $\sigma$, $h_3$ and $h_4$ \citep{Marel_Franx_93}.
Multiplicative Legendre polynomials are also included to take into account
possible internal dust reddening and residual spectrum slope variations due to
the errors in assumed instrument spectral response.  Ionized-gas
emission lines and traces of the subtracted strong airglow lines do not affect
the solution due to masking of the narrow 20-\AA-wide regions around them.
Moreover, excluding age-sensitive Balmer lines from the fit neither biases age
estimates nor degrades significantly the quality of the age determination (see
details in \citet{virgo_hbeta_sensitivity} and Appendix A2 in
\citet{nburst_a}). In order to achieve the required signal-to-noise ratio per
spatial bin we performed adaptive binning of the long-slit spectra as well as
of the integral-field spectra by using Voronoi-tessellation scheme
\citep{voronoi_binning}.  Fig.~\ref{ic719_kin_nbursts} shows the parameters
of the stellar kinematics for IC~719 derived in such a way. 

The uncertainties of the parameters derived were returned by the
minimization procedure using flux errors propagation through all data reduction steps.
\citet{nburst_a} performed some Monte Carlo simulations for the 
{\sc nbursts} fitting technique and demonstrated the
consistency between the uncertainties of the parameters returned by the
minimization procedure and the real error distributions.

By applying this approach, we found that the stellar velocity dispersion demonstrates 
two off-centred maxima and high absolute values of $h_3$, $h_4$, up to 2.5 (see
Fig.~\ref{ic719_kin_nbursts}). Such values of $h_3$, $h_4$ correspond to a
strongly asymmetric stellar LOSVD. In order to analyze in detail the stellar
LOSVD we probed then a non-parametric recovery technique, which does not
require a priori knowledge of the LOSVD shape.

\subsection{Non-parametric LOSVD} 

The only assumption we used is that an observed galaxy spectrum logarithmically
rebinned in the wavelength domain can be represented by convolution of a
velocity distribution $\mathcal{L}$ and a typical stellar (template) spectrum.
We used the SSP model from the {\sc nbursts} fitting, preconvolved with the LSF
output, as a template spectrum. The convolution can be considered as a linear
inverse problem whose solution (vector $\mathcal{L}$) can be estimated by the
least squares method. The solution is very sensitive to the noise in the data.
Hence we chose to regularize the problem be requiring the LOSVD to be smooth.
In order to do so, we choose the smoothing regularizing matrix operator in the
following form:
$\mathcal{P}(\mathcal{L})=\mathcal{L}^T\cdot\mathcal{D}_i^T\cdot\mathcal{D}_i\cdot\mathcal{L}$,
where matrix $\mathcal{D}_i$ - is the $i^{th}$ order difference operator. For
details, see chapter 19 of ``Numerical Recipes'' \citep{numrecipes}.  The
regularized inverse problem, where the function to be minimized is given by
$\Omega = \chi^2 + \lambda \mathcal{P}(\mathcal{L})$, can be expressed as a
linear matrix equation and solved by the BVLS algorithm (\citet{bvls}) which
allows to constrain the solution by positive values.  We performed numerous
tests by varying the order of the difference operator and penalization
coefficient $\lambda$, and concluded that the optimal values are $i=5$ and
$\lambda = 0.01 \lambda_e$. The choice of $\lambda$ equal to $\lambda_e$ will
tend to make both parts of the minimization function $\Omega$ to have
comparable weights. The technique used here is similar to that developped by
\citet{UGD_method}. The stellar LOSVD derived from the long-slit data for
IC~719 is presented in Fig.~\ref{PV_diag} as a position-velocity diagram. The
Gauss-Hermite representation of the stellar LOS recovered under one-component
approach also is shown at Fig.~\ref{PV_diag}. One can see that there
is no acceptable agreement between the non-parametric approach and the
Gauss-Hermite representation. Therefore, the Gauss-Hermite parametrization
cannot be applied to the complex LOSVD where two comparable peaks have a
significant velocity separation.

\begin{figure*}[htbp]
\epsscale{1} 
\centering
\includegraphics[width=1\textwidth]{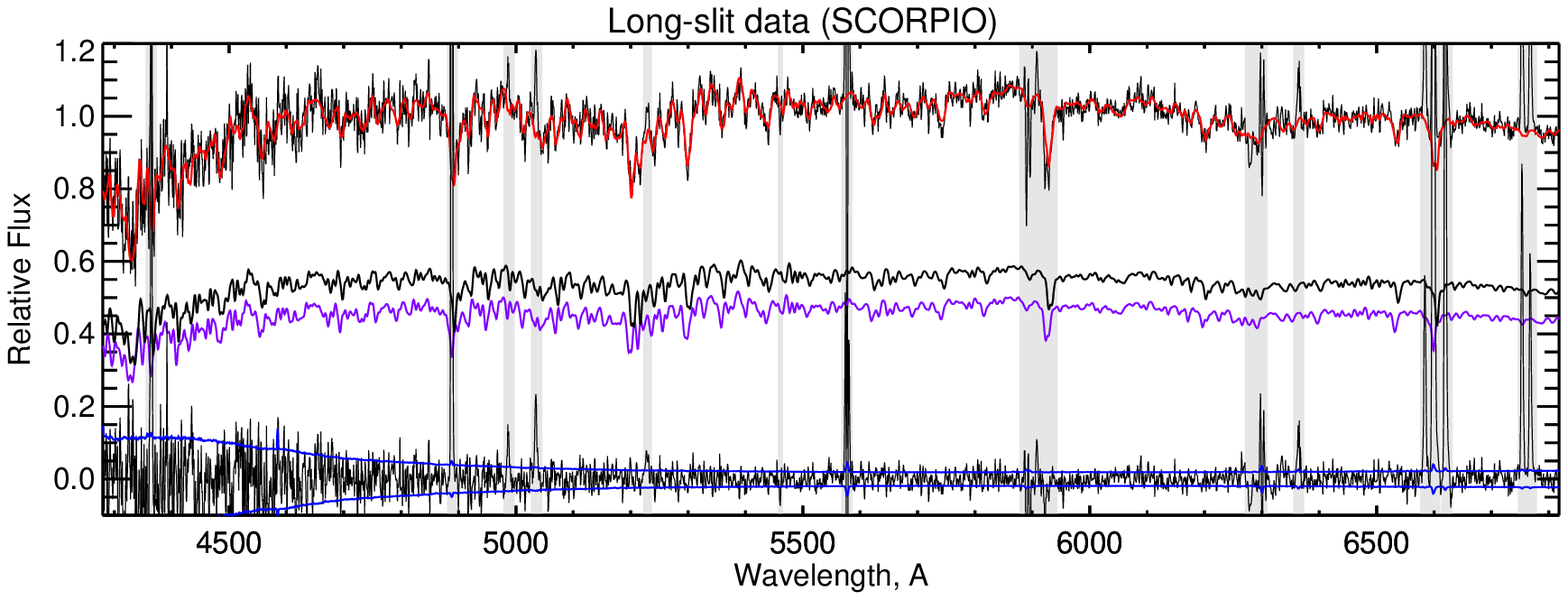}
\includegraphics[width=0.7\textwidth]{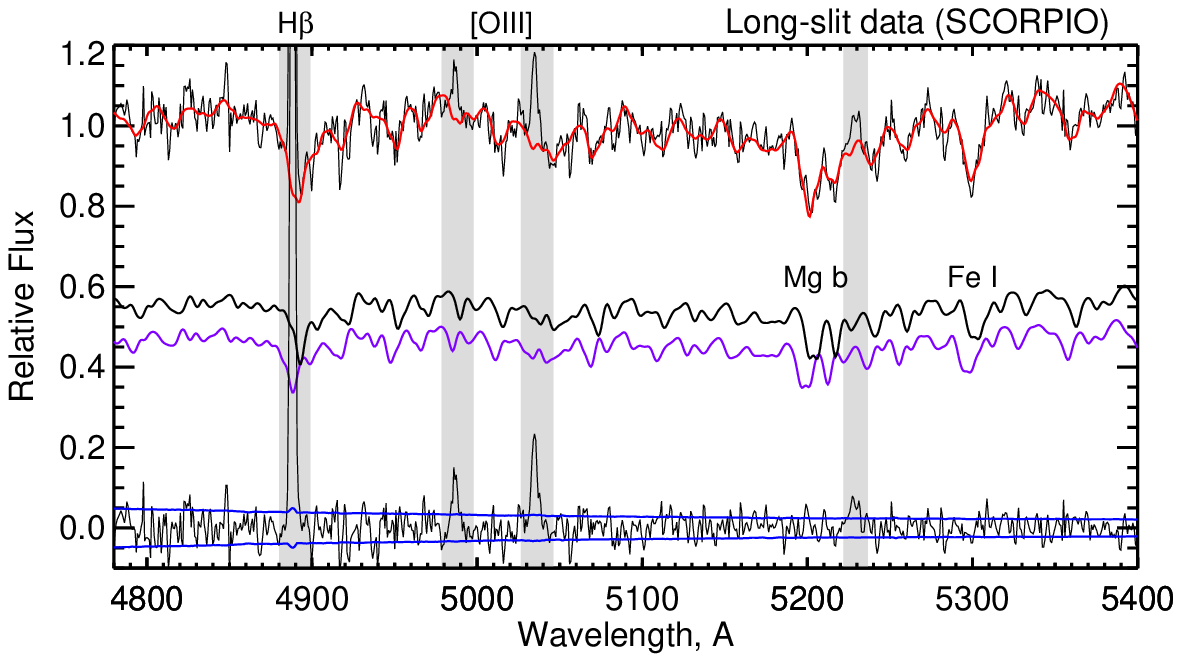}
\includegraphics[width=0.7\textwidth]{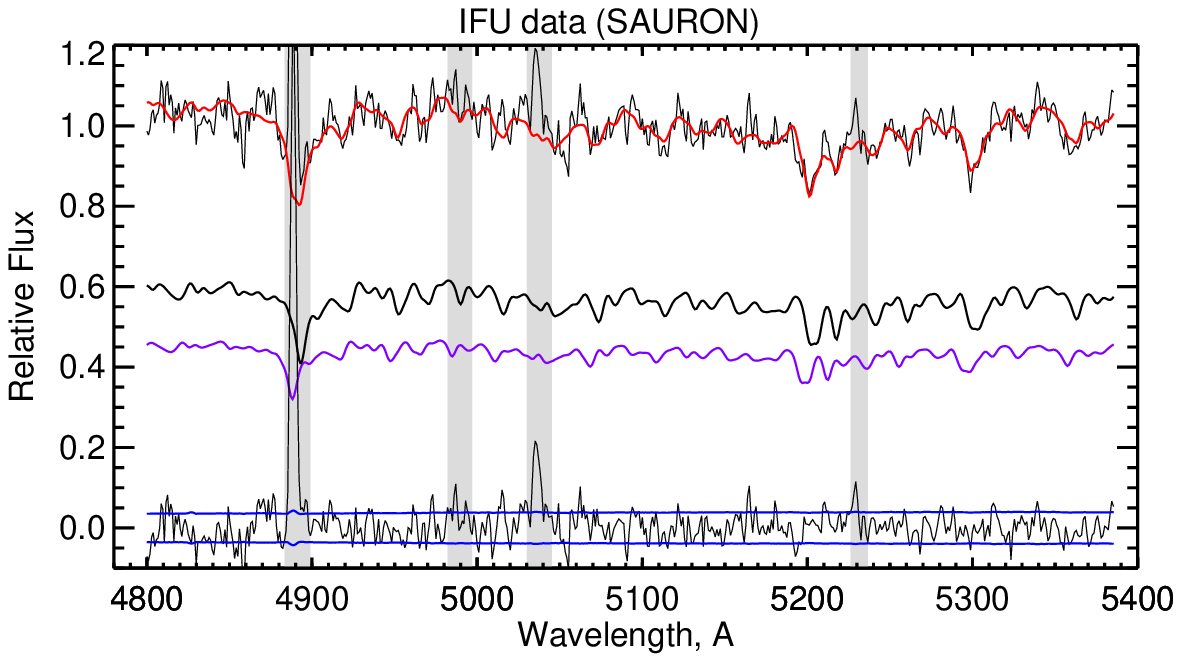}
\caption{The galaxy spectrum in the spatial bin at 7-8 arcsec south-west from
the galaxy centre extracted from the long-slit (top and middle panels) and IFU
dataset (bottom panel). The data are shown in black, the red line shows
best-fit model which is the sum of the two stellar components (purple and
black). Residuals and noise level are also shown. The grey vertical stripes
correspond to masked narrow regions around emission lines which are excluded
from fitting procedure.}
\label{spectral_decomposition} 
\end{figure*}

\subsection{Two-component fitting}

Fig.~\ref{PV_diag} demonstrates that the stellar LOSVD for IC~719 has a
complex two-peaked structure. We have decomposed the stellar LOSVD derived by
the non-parametric technique into two separate gaussians in order to evaluate
line-of-sight velocity and velocity dispersion of every component.  The centers
of the gaussians are overplotted at the position-velocity diagram as black and
blue circles (Fig.~\ref{PV_diag}).

\begin{figure*}[p] 
\centerline{
\includegraphics[width=0.3\textwidth]{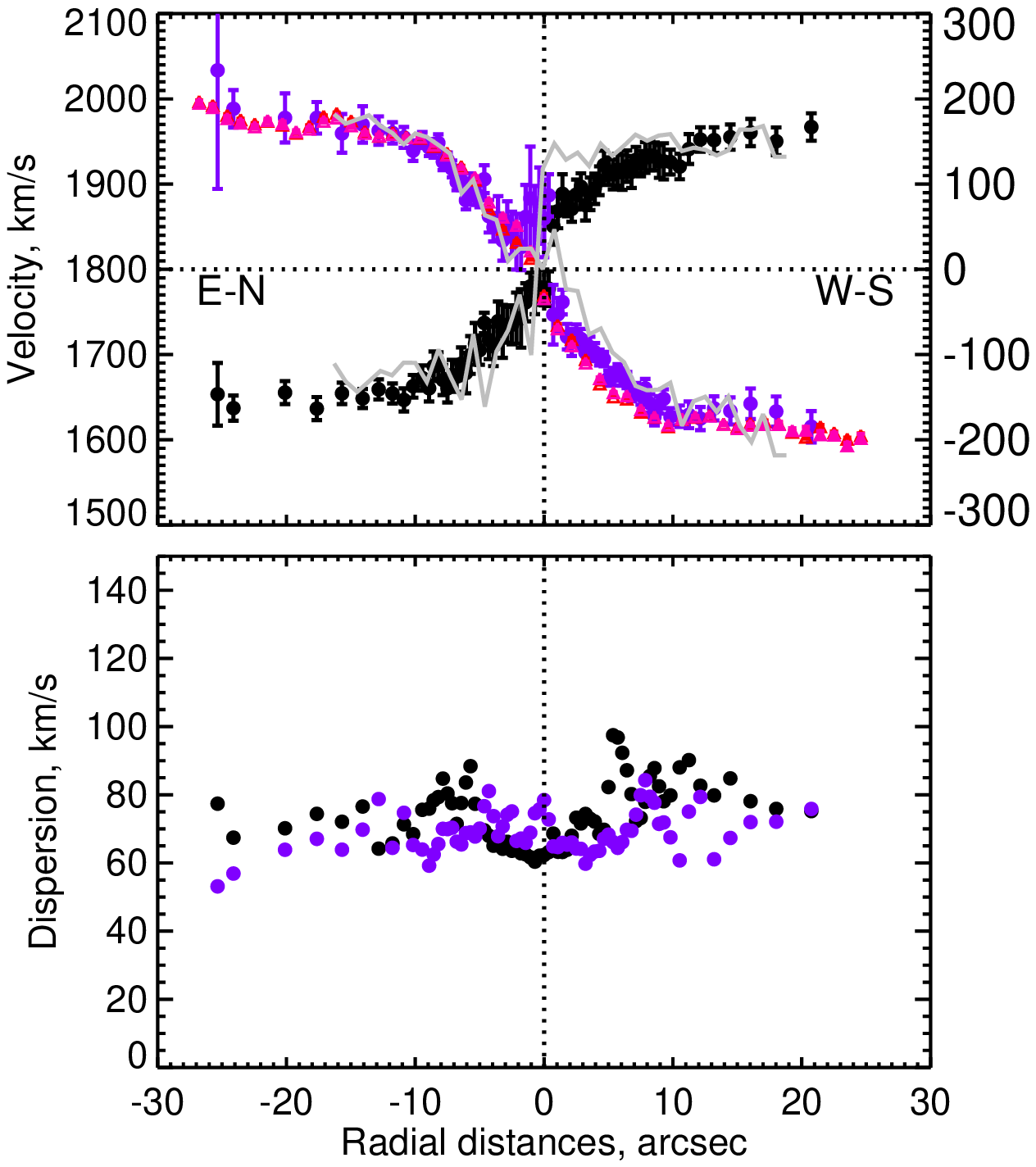}
\includegraphics[width=0.3\textwidth]{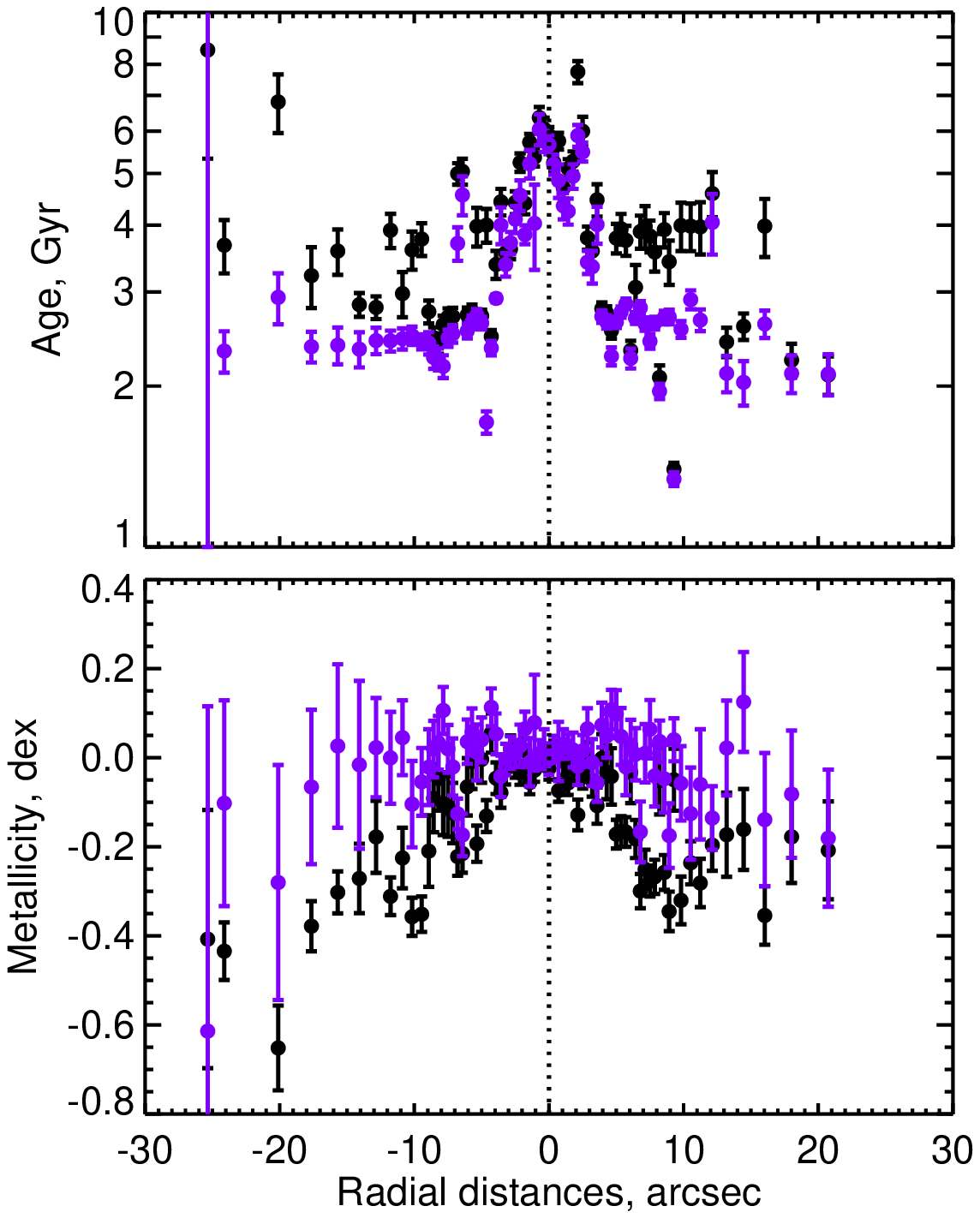}
\includegraphics[width=0.3\textwidth]{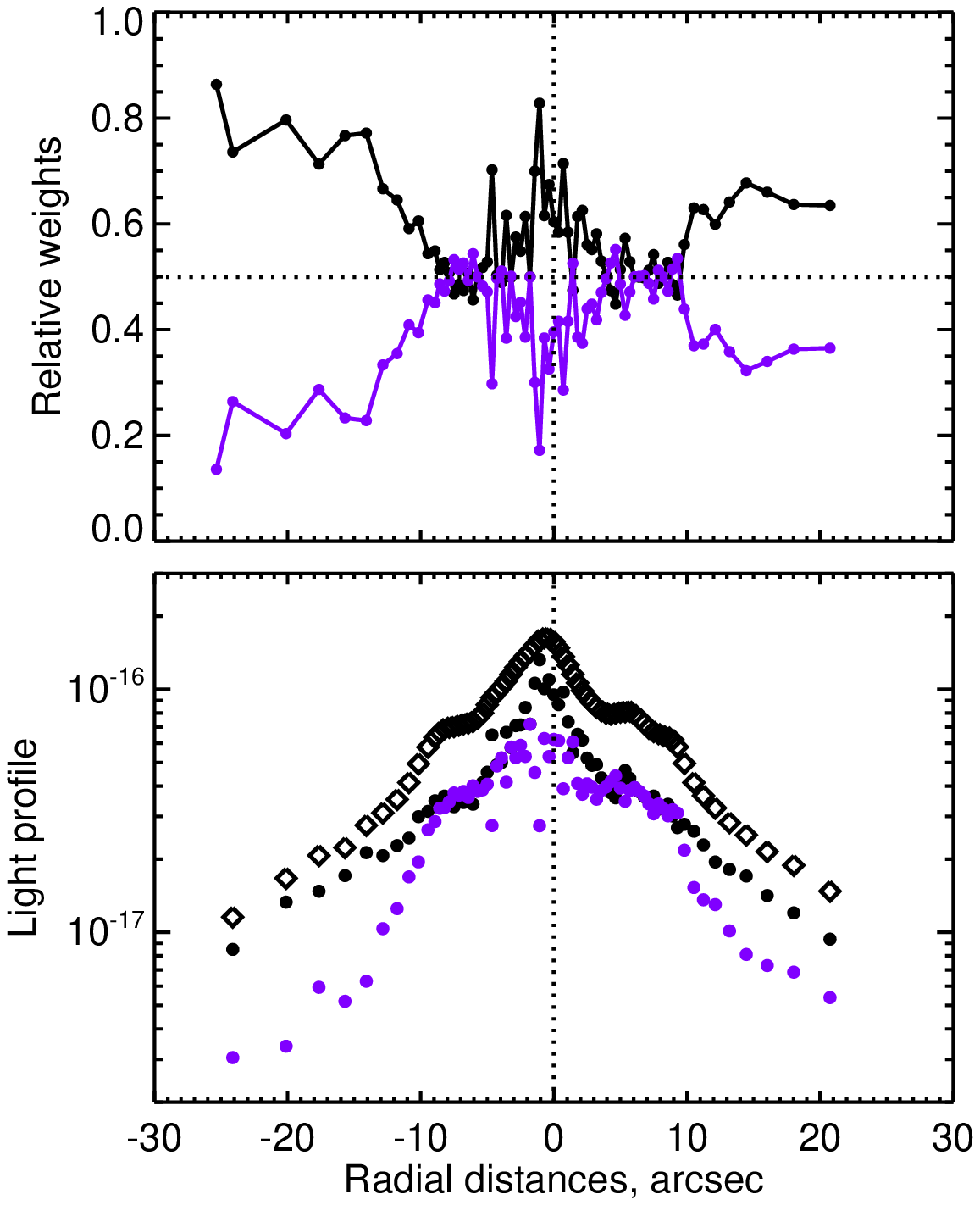}}
\caption{The results of the two-component analysis of the long-slit spectrum
for IC~719: left top -- LOS velocity profiles, both for the stars (black and
blue dots) and for the ionized gas (pink triangles for the
[N\ii]$\lambda$6583); left bottom -- stellar velocity dispersion profiles, mid
top -- SSP-equivalent age profiles for two stellar components, mid bottom --
metallicity profiles for two stellar components; right top -- relative
contribution of components, right bottom -- open diamonds shows light profile
which was obtained by integration spectra along long-slit, close symbols
correspond to light profiles for each component.  The binning is made along the
slit to maintain $S/N=20$ per pixel. The grey lines at left panel
correspond to SAURON data extracted along major axis. }
\label{long_slit_decomposed_profiles} 

\centerline{
\includegraphics[width=0.8\textwidth]{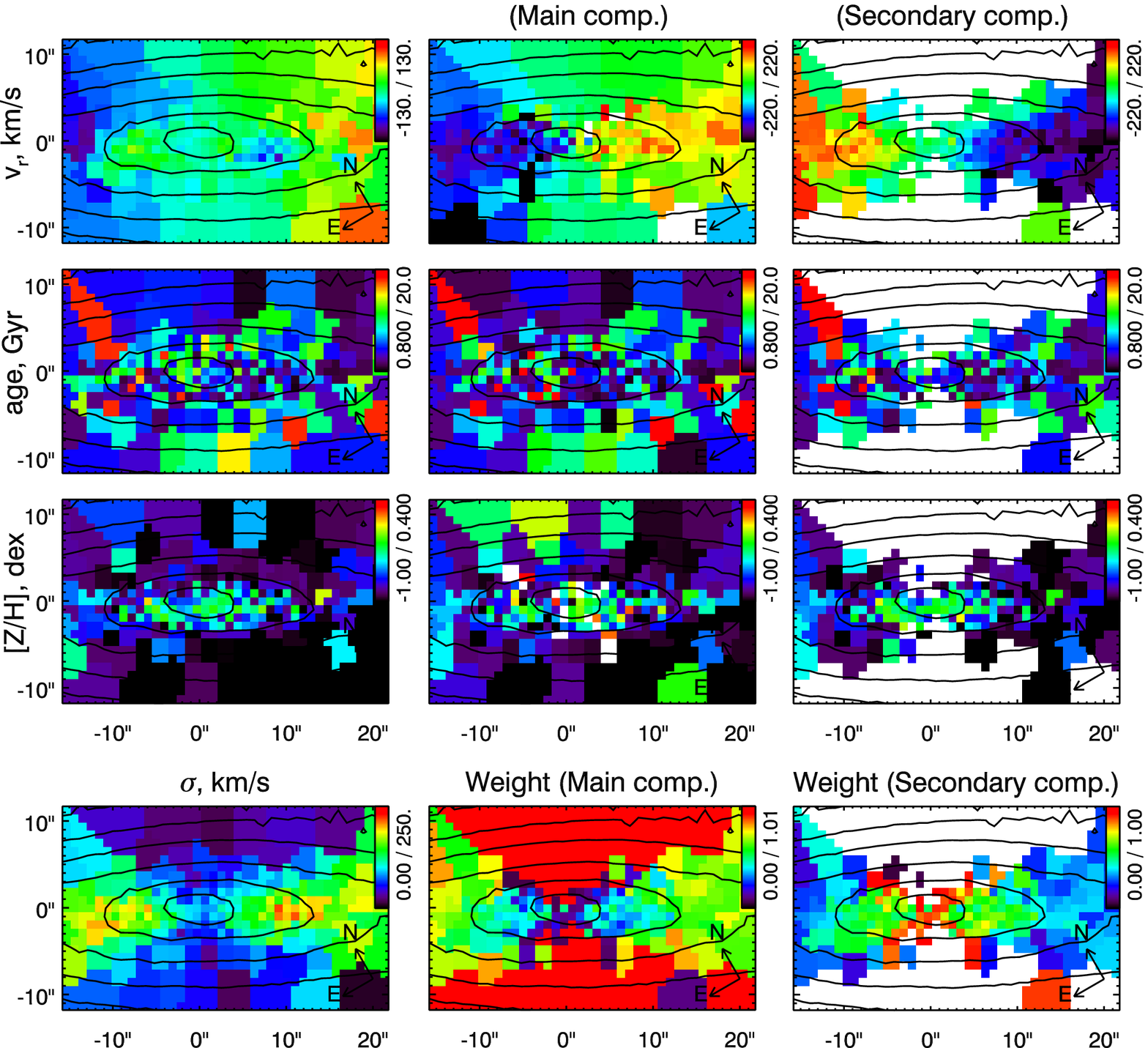}}
\caption{The results of the analysis of the SAURON spectra for IC 719: left
column shows the results of fitting by a single stellar component, and the
central and right columns show the two-component fitting results. From top to
bottom the LOS velocity field, the SSP-equivalent age map, the metallicity map
are presented. The bottom row consists of stellar velocity dispersion field for
one-component stellar fitting (left column) and the maps of relative
contribution of every stellar components (mid and right columns). Note that
color bar has a different range for the velocity fields at left column and
other columns.}
\label{sauron_decomposed_maps} 
\end{figure*}

In order to study the stellar populations of the counter-rotating disks we
separated their contributions into the observed spectrum by using an advanced
implementation of the {\sc nbursts} full spectral fitting technique. In
general, this approach is similar to that described in
section~\ref{nburst_description} but now we use a two-component model where the
optimal template consists of a linear combination of two SSPs characterized by
their ages and metallicities, each of them convolved with its own
Gaussian-shaped LOSVD. The relative luminosity contribution of one of the
components, $p$, is included into the parameter set of the minimization loop.
The contribution of the other component is equal to $1-p$. The multiplicative
polynomial continuum factor is similar for both components. By undertaking
the re-analysis of the long-slit data, we have decided to keep the velocity dispersions 
equal to the values, which were derived from the decomposition of the non-parametric 
LOSVD into two Gaussians. The range of the stellar velocity dispersions obtained for 
the both Gaussian components is rather narrow, $\sigma_\star=60-80$ km/s, so during 
the IFU data analysis we fixed velocity dispersions of the components to be equal
to 70 km/s. It is why there is no error bars at the velocity dispersion
profiles in Fig.~\ref{long_slit_decomposed_profiles} and there are no velocity
dispersion maps in Fig.~\ref{sauron_decomposed_maps}. If we take the
dispersions of both components as free parameters, the velocity profiles and
profiles of the stellar population properties look more noisy due to the
additional degree of freedom. 

In Fig.~\ref{spectral_decomposition} we demonstrate the quality of our spectral
fitting results by plotting the observed and model spectra in the same,
arbitrarily taken, spatial bin for two datasets -- our long-slit data and the
SAURON IFU data. The kinematical and stellar population parameters of the
stellar components separated in such a way as well as the relative luminosity
contributions and light profile decomposition results are shown in
Fig.~\ref{long_slit_decomposed_profiles} for the long-slit data and in
Fig.~\ref{sauron_decomposed_maps} for the SAURON data. We have succeeded to
extract two stellar components counterrotating with the similar velocities
about $\sim 200 \cdot \sin i$ km/s. The component which corotates the ionized
gas (see below) gives lower contribution to the disk surface brightness at
$R>10\arcsec$ than its counterpart, so we would treat is as a `secondary'
stellar component; in general it looks somewhat younger ($T_{sec}\approx2.5$
Gyr) and more metal rich ([Z/H]$_{sec}\approx -0.1 \div 0.0$ dex) than the
other one ($T_{main}\approx4$ Gyr, [Z/H]$_{main}\approx -0.3 \div -0.2$ dex).

Two derived counterrotating stellar components seem to be dynamically cold disk
structures, with their ratio of the maximum rotation velocity to velocity dispersion
$V_{rot}/\sigma \approx 3$ for each component. The identification of the large-scale
stellar structures in IC~719 as the cold flat disks is supported by surface
photometry. We extracted azimuthally averaged surface brightness
profiles in the SDSS $g,r,i$-bands and in the UKIDSS $K$-band using the IRAF
task {\sc ellipse} \citep{ellipse} (see Fig.~\ref{view_prof}) and found that the
profiles are quasi exponential with a compact nuclear concentration and no
presence of a massive bulge. At the radii of $r=7''-10''$ one can see an excess of the
surface brightness, which apparently corresponds to the concentration of the secondary 
stellar component. This luminosity excess looks like a ring structure at the bivariate
picture of the residuals between $K$-band image and exponential disk model
constructed over the outer regions ($r>20''$) of the galaxy by using the GALFIT
package \citep{galfit}.

\subsection{Ionized gas}

An emission-line spectrum of every spatial bin was obtained by subtracting 
the stellar contribution (i.e., the best-fitting stellar population model) from the 
observed spectrum. This step provided a pure emission spectrum uncontaminated by
absorption lines of the stellar component that is especially important for the
Balmer lines.  Then we fitted emission lines with Gaussians pre-convolved with
the instrumental LSF in order to determine the line-of-sight (LOS) velocities
of the ionized gas and emission-line fluxes.  The LOS velocity profiles and
maps of velocities and fluxes of the emission lines are shown in
Fig.~\ref{long_slit_decomposed_profiles} and Fig.~\ref{sau_emissions},
respectively.

The line fluxes have been corrected for the internal interstellar extinction as
well as for the Galactic extinction according to \citet{schlegel98}. The color
excess E(B-V) corresponding to the internal dust reddening was determined from
Balmer decrement using the theoretical line ratios $F(H\alpha)/F(H\beta)=2.87$
for the electron temperature $T_e=10000$ K \citep{Osterbrock_Ferland_2006} and
the parametrized extinction curve \citep{fitzpatrick_1999_extinction_curve}.

We plotted our measurements of emission lines onto the classical
excitation-type diagnostic diagrams to identify the gas ionization source
(Fig.~\ref{BPTdiag}). One can see that all measurements are located in the
area corresponding to photoionization by hot stars except a few central points
which correspond to shock-like excitation or to that powered by a weak (LINER)
AGN.  

Due to the wide spectral range of the SCORPIO long-slit spectra covering a set
of strong emission lines (\Hb, [O\iii]$\lambda$5007, \Hb,
[N\ii]$\lambda6548,\lambda6583$, [S\ii]$\lambda6717,\lambda6730$) we are able
to determine oxygen and nitrogen abundances of the ionized gas. To do this we
used so-called NS-calibration method \citep{pilyugin_ns_calib11} which does not
require [O\ii]$\lambda3727+\lambda3729$ lines. This NS calibration uses strong
emission lines O$^{++}$, N$^+$, S$^+$ and is based on the spectra of H\ii\,
regions with measured electron temperatures as a calibration data set. In order
to compare oxygen and nitrogen abundances of the ionized gas with the
metallicity of the stellar component, we plot all the radial profiles of the
abundances in Fig.~\ref{gas_abund}.

\section{Discussion}

The LOSVD of IC~719 can be easily splitted by eye into two counterrotating
stellar components (Fig.~\ref{PV_diag}). So for this galaxy we have undertaken
the spectrum fitting by two SSPs with different kinematical and stellar
population parameters.

\begin{figure*}[h!t] 

\centering
\includegraphics[width=0.3\textwidth]{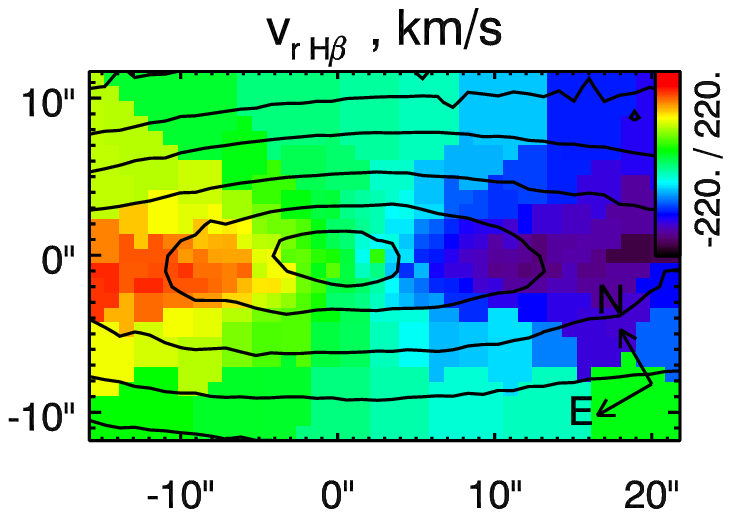}
\includegraphics[width=0.3\textwidth]{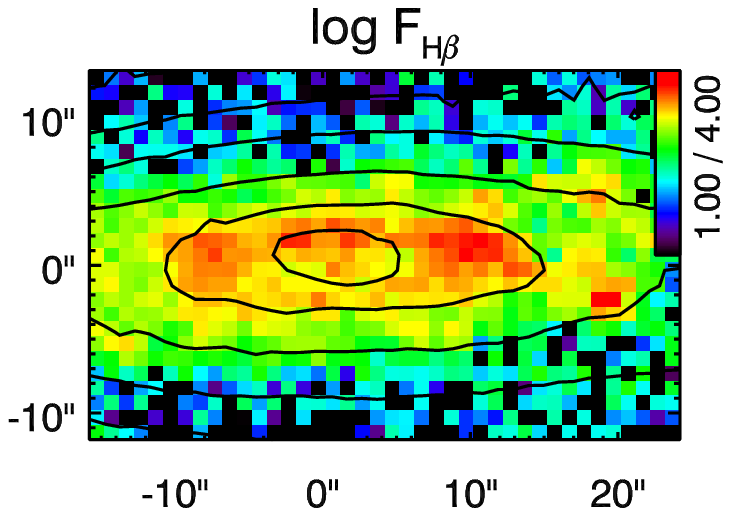}
\includegraphics[width=0.3\textwidth]{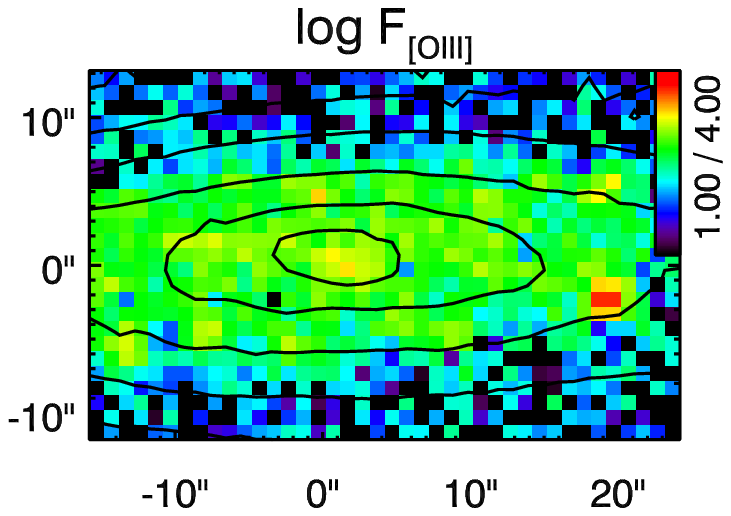}
\caption{The results of the emission-line component analysis by using the
SAURON data: left -- the ionized gas velocity field constructed using H$\beta$
line, mid -- the intensity map for the emission line H$\beta$, right -- the
intensity map for the emission line [O\iii]$\lambda$5007}
\label{sau_emissions}

\centerline{
\includegraphics[width=0.32\textwidth]{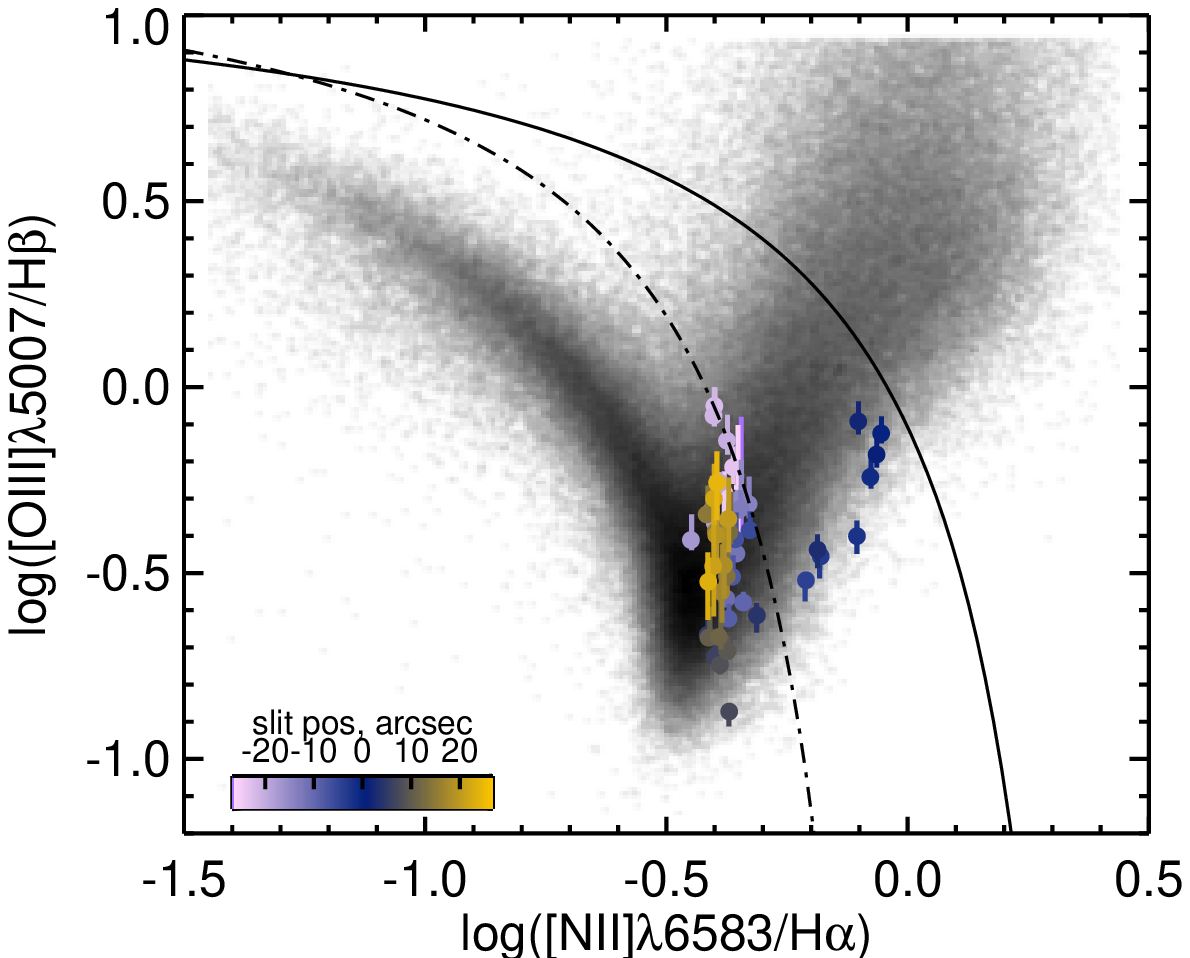}
\includegraphics[width=0.32\textwidth]{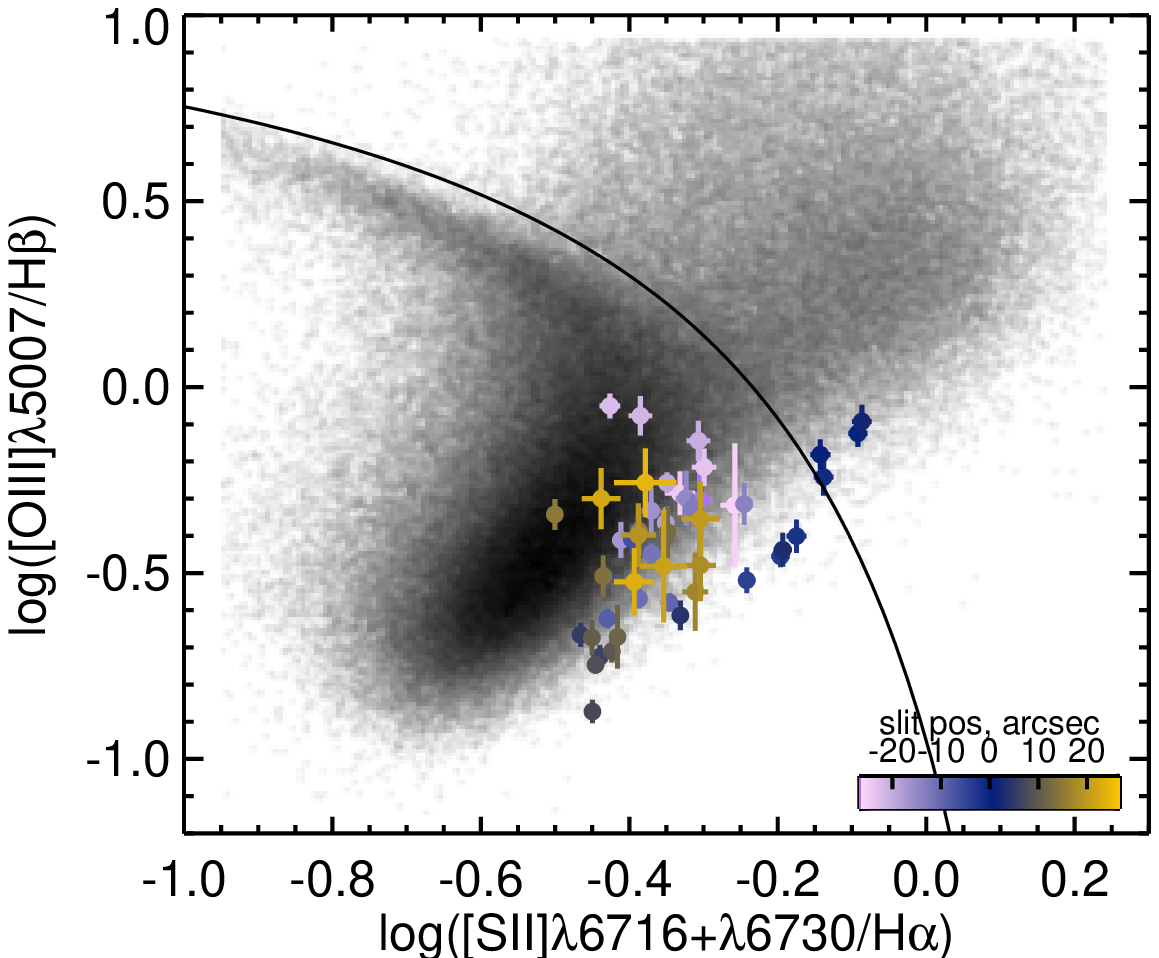}
\includegraphics[width=0.32\textwidth]{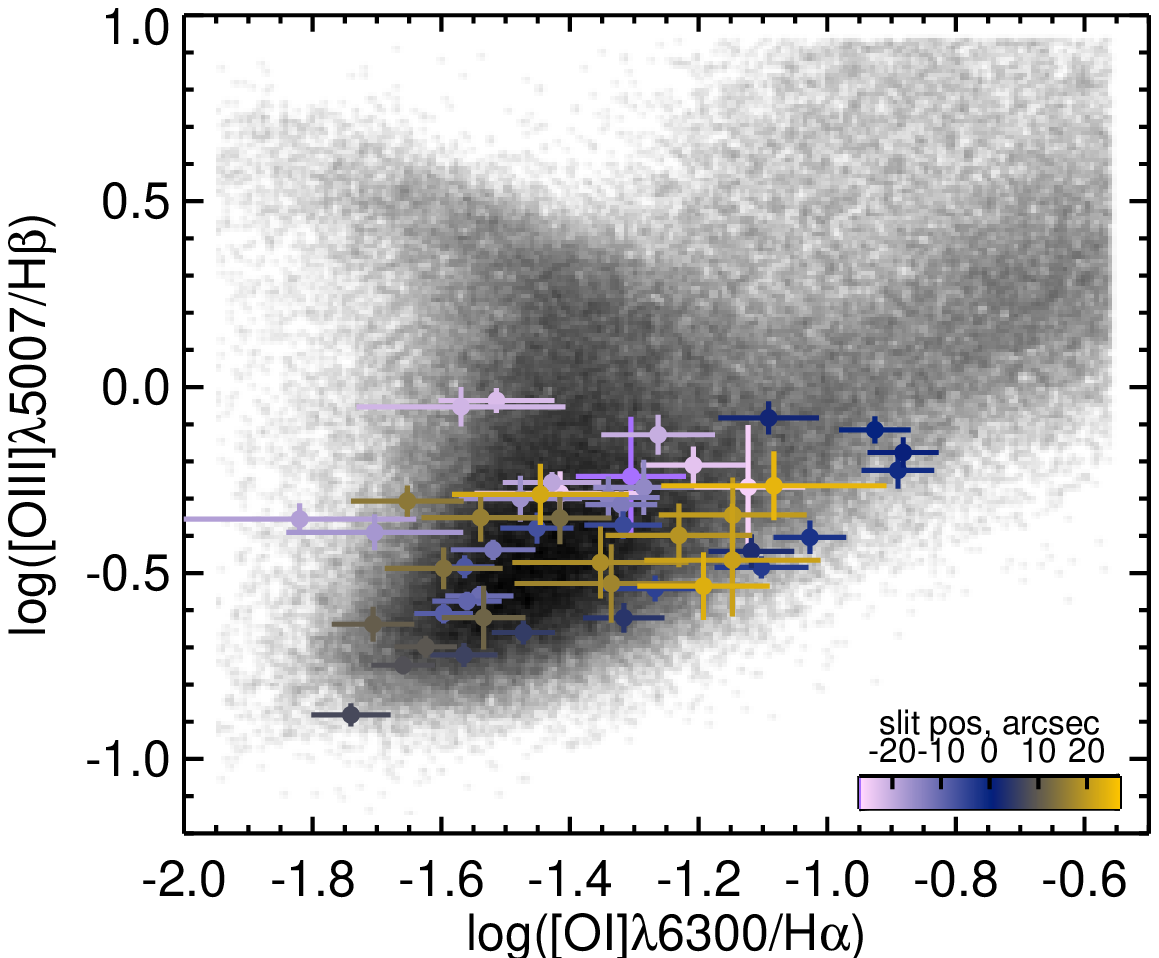}}
\caption{Excitation diagnostic diagrams comparing the emission-line ratios:
[N\ii]/\Ha  vs. [O\iii]/\Hb (left), [S\ii]/\Ha  vs. [O\iii]/\Hb (mid),
[O\i]/\Ha vs.  [O\iii]/\Hb (right). The points show the extracted emission line
ratios from the long-slit spectra for IC~719. The color of points corresponds
to the distance from the galaxy center. Distribution of measurements
of the line ratios for galaxies from SDSS survey with high signal-to-noise
ratio ($S/N>3$ in every line) is showed by grey color. The black curves, which
separate the areas with the AGN/LINER excitations from areas with the
star-formation-induced excitation, are taken from \citet{kauffmann2003}
(dash-dot curve) and from \citet{kewley06} (solid curve). }
\label{BPTdiag}
\end{figure*}

The early analysis by the ATLAS-3D team that was using the kinemetry approach
involving one-component stellar population with the unified kinematics,
returned the classification of IC~719 as a rotator, slow in the inner part. The
galaxy was also included into a so called `2$\sigma$' type, so was treated as a
galaxy having two off-centered maxima of the stellar velocity dispersion
\citep{atlas3d_2, atlas3d_3}.  In fact, when the correct approach including two
kinematically (and evolutionarily) decoupled SSPs is applied, we find two
fast (counter-) rotating stellar components even in the innermost region
(Fig.~\ref{sauron_decomposed_maps}).  The off-centered spots of visibly high
stellar velocity dispersion (or of a widened LOSVD) are in fact the locations
the nearest to the center where two LOSVD branches start to separate. The
gaseous component extracted from the SAURON data after the fitting  by two SSP
rotates just as one of the stellar components; the surface intensity of the
H$\beta$ emission demonstrates a ring-like distribution with the radius of
about $10\arcsec$ (Fig.~\ref{sau_emissions}). 

\begin{figure}[htbp]
\centering
\includegraphics[width=0.45\textwidth]{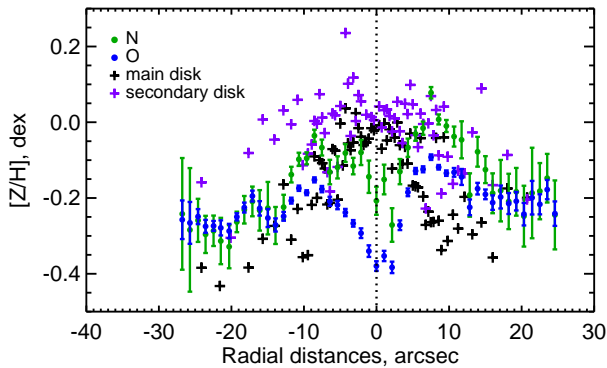}
\caption{Radial profiles of gaseous (filled circles) and stellar abundances
(crosses). Green and blue circles with error bars correspond to nitrogen and
oxygen abundances with subtracting solar value, respectively. We adopted a
solar nitrogen abundance $12+\log N/H$ to be 7.83 then the solar oxygen abundance $12+\log
O/H$ is 8.69 \citep{asplund_solar_abund}. The abundances in the nucleus are
unreliable due to different gas excitation mechanism there. }
\label{gas_abund}
\end{figure}

By applying our two-component fitting to the IC~719 long-slit data obtained
with the SCORPIO (Fig.~\ref{long_slit_decomposed_profiles}), we have obtained
two counterrotating stellar components once again; one of the LOS stellar
velocity profiles coincides perfectly with the gas velocity profile, the other
is mirroring. The amplitudes of both stellar LOS velocity profiles are similar.
Beyond the central part of the galaxy, the stars corotating the ionized gas
look somewhat younger ($T_{sec}\approx2.5$ Gyr) and more metal-rich
([Z/H]$_{sec}\approx -0.1 \div 0.0$ dex) than the counterrotating component
($T_{main}\approx4$ Gyr, [Z/H]$_{main}\approx -0.3 \div -0.2$ dex). The BPT
diagram diagnostics of the pure emission-line spectrum, which was obtained by
subtracting the model stellar population spectrum from the observed one, has
shown (Fig.~\ref{BPTdiag}) the following picture: the gas excitation in the
nucleus is shock-like or powered by a weak (LINER) AGN, but in the ring and
beyond  it the gas is ionized by young stars. After measuring the gas abundances
by using the emission line flux ratios and NS calibration relation from
\citet{pilyugin_ns_calib11}, we find that in general the ionized gas
metallicity is higher then the metallicity of the main stellar component and
tends to be smaller or equal to the metallicity of the secondary component (see
Fig.~\ref{gas_abund}).

For the first time the technique of spectroscopic decomposition was presented
by \citet{coccato_2011_n5719}. They simultaneously measured both the kinematics
and stellar population properties of the counter-rotating stellar disks of
NGC~5719 and found that the stellar component, which rotates in the same
direction as the ionized gas, is younger, less rich in metals, more
$\alpha$-enhanced, and less luminous than the main galaxy disk. Their
investigation defiantly confirmed the gas accretion scenario for the secondary
stellar disk of NGC~5719. Later this team studied stellar populations in the
two more well-known galaxies NGC~3593 and NGC~4550 hosting counter-rotating
disks \citep{coccato_2012_n3593_n4550}. In the both galaxies the secondary
stellar components are less massive, more metal poor and $\alpha$-enhanced than
the main galaxy stellar disks, and co-rotate the ionized gas. These findings
rule out an internal origin of the secondary stellar component and favor a
gas-accretion scenario. NGC~4550 was also studied by \citet{johnston_n4550}.
They obtained that the age of the secondary disk is a bit younger than it was
evaluated by \citet{coccato_2012_n3593_n4550}. Both papers concluded that the
most likely formation mechanism of the secondary stellar component in NGC~4550
is unusual gas accretion history.

What can be the origin of the counterrotating gas in IC~719? A standard answer
could be `merging' or `interaction'.  The chemistry of the gas which is not
very metal-poor (at least within the stellar body of the galaxy) excludes the
version that the extended counterrotating gaseous disk consists of the relic
baryons never processed through the stars, so there is no reason to suspect
cold accretion from large-scale cosmological filaments.  Meantime, IC~719
constitutes an isolated non-interacting pair with IC~718 -- a late-type galaxy
of similar luminosity.  Moreover, neutral hydrogen mapping reveals huge gas
cloud embedding both galaxies \citep{alfalfa2,atlas3d_13}. The HI extended disk
of IC~719 looks coplanar to the stellar  disk of IC~719 though is much more
extended -- up to 100 kpc from the galactic center. Is it an interaction with
gas flows directed from the late-type neighbour onto the early-type one?
Merging with some third galaxy seems improbable: the high mass of the gas (at
least 0.7 billion solar masses), together with the modest stellar mass of the
galaxy, $6 \cdot 10^{10}$ solar masses (see $M_K$ in the Table~1), implies the
necessity of merging galaxies with the mass ratio larger than 1:10; it means
that the thin stellar disk could not survive during such merging
\citep{walker,thakarryden} unless very specific conditions were provided --
intense hot halo gas cooling after the merger \citep{moster} (meantime, X-ray
haloes are not reported around these galaxies) or strictly coplanar satellite
orbiting before the merger. Meantime both stellar disks of IC~719 are rather cold
dynamically.

However, it is evident that the inner ionized-gas component having the same
angular momentum has a strong connection with the secondary stellar disk,
though star formation in the disk have started perhaps recently, because the
bulk stellar populations in the large-scale stellar disk of the galaxy are of
intermediate age, $>2$~Gyr. One can consider two accretion scenarios of the gas
onto the galaxy. In the frame of the first one, the gas was acquired
through a single accretion event but then two starbursts have happened. Early
starburst has given rise to formation of the secondary stellar component and
the last one continues now. The second scenario consists of two accretion
events, each with a subsequent starburst. We prefer the latter scenario of gas
accretion because the former one is inconsistent with the gas metallicity
measured by us. Indeed, under the first scenario we expect that the metallicity
of the self-enriched gas would be higher then the stellar metallicity of the
secondary component but the Fig.~\ref{gas_abund} shows the opposite relation.
The second scenario allows that the metallicity of the mixture of the
metal-poor external gas with the inner enriched gaseous fraction may be below
the metallicity of the secondary stellar component, that is consistent with our
measurements.

\section{Conclusions}

By adding our long-slit spectral observations to the panoramic spectroscopy
with the SAURON, we have detected extended counterrotating gaseous disk as well
as a secondary stellar component corotating the ionized gas in the lenticular
galaxy IC~719. The gas counterrotation can be traced up to the optical borders
of the galaxies; moreover, the ionized-gas velocity profiles obtained by us are
more extended than the stellar velocity profiles extracted from the same
spectral data. The gas emission-line surface intensity demonstrates ring-like
distributions within the disk of the galaxy; according to the BPT-diagnostics
from the line ratio confrontation, the gas in the rings is excited by young
stars, so the extended disk of the IC~719 possess rather intense current star
formation. We have demonstrated that the accretion history of the external gas
onto IC~719 consist of two events, each with a subsequent starburst.

In any case, this field lenticular galaxy is growing its large-scale
counterrotating stellar disk just now promising to become in the nearest future
an analog of NGC~4550 -- the S0 galaxy which has two equal-mass counterrotating
stellar disks \citep{rubin,rix} and no current star formation \citep{crockeretal}.

\acknowledgments

This research is partly based on data obtained from the Isaac Newton Group
Archive which is maintained as part of the CASU Astronomical Data Centre at the
Institute of Astronomy, Cambridge.  We have made use of the NASA/IPAC
Extragalactic Database (NED) which is operated by the Jet Propulsion
Laboratory, California Institute of Technology, under contract with the
National Aeronautics and Space Administration. In this study, we used the
UKIDSS DR8 image data available through the WFCAM science archive and SDSS DR9
data.  Funding for the SDSS and SDSS-II has been provided by the Alfred P.
Sloan Foundation, the Participating Institutions, the National Science
Foundation, the U.S. Department of Energy, the National Aeronautics and Space
Administration, the Japanese Monbukagakusho, the Max Planck Society, and the
Higher Education Funding Council for England. The SDSS Web Site is
http://www.sdss.org/. We wish to thank the anonymous referee for his/her useful
comments and suggestions. We acknowledge the usage of the HyperLeda database.
This work was supported by the Ministry of Education and Science of the Russian
Federation and Russian Foundation for Basic Research (projects
no.~10-02-00062a, no.~12-02-00685а) and Presidential Grant no.~MD-3288.2012.2.
IYK is also grateful to Dmitry Zimin's non-profit Dynasty Foundation.

\clearpage

\end{document}